\documentclass[twocolumn,prc,nofootinbib,showpacs]{revtex4}
\usepackage{graphicx}
\newcommand{\bdec}{$\beta$-decay }
\newcommand{\be}{\begin{equation}}
\newcommand{\ee}{\end{equation}}
\newcommand{\iso}[2]{{\ensuremath{{}^{#2}\mathrm{#1}}}}

\begin{document}
\flushbottom
\title{Experimental study of ${\mathbf{{}^\mathbf{113}}\mathbf{Cd}}$  beta decay using CdZnTe 
detectors}
\author{C. Goe{\ss}ling$^{a}$, M. Junker$^{b}$, H. Kiel$^{a}$, D. Muenstermann$^{a}$, S. Oehl$^{a}$, K. Zuber$^{c,d}$}
\address{$^{a}$Lehrstuhl f\"ur Experimentelle Physik IV,
Universit\"at Dortmund,\\ Otto--Hahn Str.~4,
44227 Dortmund, Germany\\
$^{b}$Laboratori Nazionali del Gran Sasso, Assergi, Italy\\
$^{c}$Denys Wilkinson Laboratory, University of Oxford, Keble Road, Oxford OX1 3RH\\
$^{d}$Dept.~of Physics and Astronomy, University of Sussex, Falmer, Brighton BN1 9QH, UK}
\begin{abstract}
A search for the 4-fold forbidden beta decay of \iso{Cd}{113} has been performed with CdZnTe 
semiconductors. With 0.86 kg $\cdot$ d of statistics a half-life for the decay of $T_{1/2} = (8.2 \pm 0.2 
(stat.) ^{+0.2}_{-1.0} (sys.)) \nolinebreak \cdot 10^{15}$yrs has been obtained. This is in good agreement with published values. 
A comparison of 
the spectral shape with the one given on the Table of Isotopes Web-page shows a severe deviation. 

\end{abstract}
\maketitle
\section{Introduction}
\begin{figure}
\centering
\includegraphics[width=6cm]{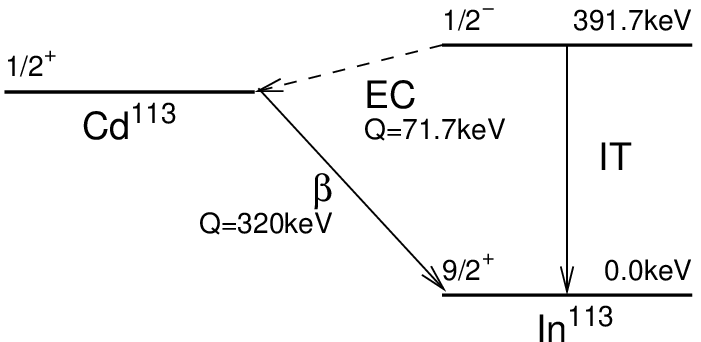}
\caption{Level scheme of the \iso{Cd}{113} decay.}
\label{levels}
\end{figure}

In past decades the investigation of \bdec played a major role
in understanding the structure of weak interactions.
Nowadays, such studies seem to be a little bit out of fashion, but
there are still a lot of interesting problems to be investigated.
Among them is the measurement of higher order transitions like
4-fold forbidden non-unique decays.
Their log ft-values are larger than 20 and correspondingly half-lives
around $10^{15}$ yrs and longer have to be measured.
Only three isotopes of this type are known, \iso{V}{50}, \iso{Cd}{113}
and \iso{In}{115}. No beta spectrum measurement is reported for \iso{V}{50}
and only two recent half-life determinations for \iso{In}{115} \cite{pfe79,bar04}. \\
The transition \iso{Cd}{113} $\rightarrow$ \iso{In}{113} is characterised as a 
$1/2^+ \rightarrow 9/2^+$ transition with a log ft-value estimated to be 23.2 \cite{sin98} 
(Fig.~\ref{levels}).
A first attempt to measure this decay with a small CdTe device
resulted in a rather large half-life range of $(4-12) \cdot 10^{15}$ yrs
as reported in \cite{mit88}. Two more recent measurements using $\mathrm{CdWO}_4$ exist. 
One measurement using the material as a scintillator resulted in a half-life 
of $(7.7 \pm 0.3) \cdot 10^{15}$ yrs \cite{dan96}.
The second experiment used $\mathrm{CdWO}_4$ as a cryogenic
bolometer and obtained a half-life of $(9.0 \pm 0.5 (stat.) \pm 1 (sys.))
\cdot 10^{15}$ yrs \cite{ale94}. 
Here, we report on a completely independent new experimental approach
by using room temperature CdZnTe semiconductor detectors. 
The measurement has been performed within the programme of test measurements for the new
COBRA double beta decay experiment \cite{zub01}.  

\section{Experimental setup}
The data presented are obtained with four 1\,cm$^3$ CdZnTe
detectors provided by eV-PRODUCTS based on coplanar grid technology,
i.e.~reading out only the electron signal. Measurements have been
performed in the Gran Sasso Underground Laboratory (LNGS) in Italy
providing a shielding of about 3500 mwe. The four bare detectors are
mounted in a copper brick with all preamplifier electronics moved
to about 25\,cm distance. This copper brick itself is part of a 
$(20\,$cm)$^3$ cube made out of electropolished copper and a further passive
shielding of 15\,cm of lead. The whole setup is located in
a Faraday cage made from copper plates. The cage is surrounded by a neutron shield,
consisting of 7\,cm thick boron-loaded polyethylene plates and 
additional 20\,cm of paraffin wax at the bottom. The full paraffin shielding
was finished at a later stage.\\
The energy resolution and stability of the detectors is calibrated regularly with the help of
\iso{Cs}{137}, \iso{Co}{60} and \iso{Th}{228} sources. For the four detectors a CAMAC based
data acquisition system
is used. The signals are fed into four 13-bit peak sensing ADCs (types LeCroy 3511 and 3512).

\section{Data extraction}
\begin{figure}
\centering
\includegraphics[width=\linewidth]{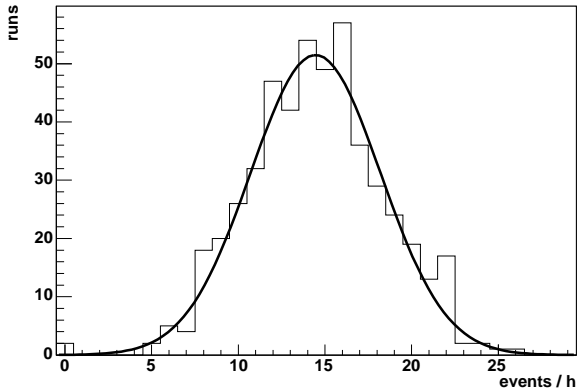}
\caption{Event rate distribution of individual runs for a selected period of time, shown for the energy range of 100 - 320 keV. 
Runs with electronic disturbances would show up with rates well beyond 30 events per hour and can be easily
rejected.}
\label{rates}
\end{figure}

The total measuring period consists of 4781 h. Each individual run 
is limited to one hour. However, some data cleaning has been necessary 
for this low energy range. 
Due to noise problems several run periods had to be performed with rather
high discriminator thresholds. 
As a consequence, one detector cannot be used for the low energy analysis
presented here. Thus 25 \% of 
the total data had to be rejected. In addition, some data sets had
to be rejected because of short term distortions of the detectors.
If the event rate per hour is histogramed, the distorted runs can easily be
identified by having exceptional high event rates (Fig.~\ref{rates}). This has been done for
every detector independently. In this way further 57 runs were discarded. An instability in the threshold leads to a loss of 12 days for one of the ADCs. 

For this purpose rather high energetic calibration
sources were used, thus the datasets have been checked independently 
for background lines in the
region of the \iso{Cd}{113} decay. It has been found that the 351.9 keV
gamma line from the \iso{U}{238} decay chain, which is located just
next to the endpoint of the beta spectrum, gives a suitable handle
on the quality of the calibration in the low energy region.
The final set of good data has been selected by checking the position
of this gamma line for each measurement period and taking its position
into account for the final calibration.
The final sample comprises a statistics of 0.86 kg$\cdot$d.

\section{Results}

\subsection{Half-life of ${\mathbf{{}^\mathbf{113}}\mathbf{Cd}}$ decay}
The spectral shape of the energy spectrum of electrons emitted 
in a 4-fold forbidden non-unique beta decay cannot be predicted by theory, 
hence we follow
the procedure described in \cite{dan96}. A model spectrum is built
using the form
\be
N(E) = F(Z,E) p (E+mc^2)(Q-E)^2 S(E)
\ee
with $F(Z,E)$ as the Fermi function, $Q$ the endpoint energy of the 
beta spectrum, $p$ the electron momentum and
$E$ the kinetic energy of the electron.
The correction function $S(E)$ is assumed to be a polynomial
of the form
\be
S(E) \sim p^6 + C_1 p^4q^2 + C_2 p^2q^4 + C_3 q^6
\ee
with $q$ as the neutrino momentum. Such a high order polynomial provides
a reasonable fit and is motivated by theoretical arguments on 4-fold forbidden 
unique beta decay. However, the decay of \iso{Cd}{113} is known to be non-unique 
and it is not obvious that this spectral shape is correct. 
The correction functions used in this analysis are based on the
measurement of the coefficients given by \cite{dan96}.

The corresponding model spectrum is folded with the detection efficiency,
calculated with a Monte Carlo simulation based on GEANT4. Due
to the short range of low energy electrons in CdZnTe the effect on the spectrum 
is marginal and corresponds to only a modest shift of the maximum
of 0.2 percent.

To get a more sophisticated energy resolution for the low energy
regime, intensive post
calibrations have been done using \iso{Eu}{152}, \iso{Ba}{133},
\iso{Co}{57}, \iso{Am}{241}, \iso{Co}{60} and \iso{Cs}{137} sources.
The measurements can be described well by a linear dependence of 
the energy resolution
as a function of energy in the range of 100 keV - 1.4 MeV. For the 
collected data the energy resolution is
extrapolated from the high energy calibrations done during the
measurements resulting in an energy resolution (FWHM) averaged over the
detectors of
\be
\Delta \mathrm{E (kev)} = 4.7\% \cdot \mathrm{E (keV)} + 28 \mathrm{keV}
\ee  

\begin{figure}
\centering
\includegraphics[width=\linewidth]{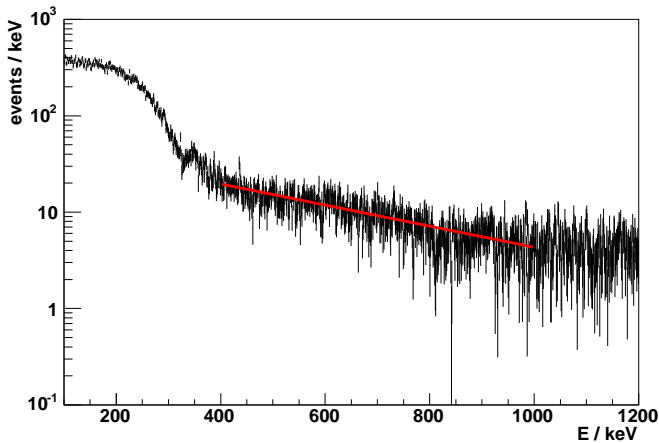}
\caption{The summed low energy spectrum of three CZT detectors.
The red curve corresponds to a exponential fit in the region 400-1000 keV.}
\label{bgfit}
\end{figure}

Before fitting the experimental data with the polynomial form,
a background component is subtracted from the single spectra.
Therefore, an exponential background of the form
\be
B(E) = B_1 \exp (-E/B_2)
\ee  
is fitted first, well motivated by the observed data (Fig.~\ref{bgfit}). To avoid any contribution of the \iso{Cd}{113} beta spectrum itself the
background fit is done using the energy range from 400-1000 keV.
The noise threshold is not included in the fit, it is well below the
chosen boundaries for the spectral fit described below.
Before performing the fit, the 351.9 keV gamma line with its Compton
spectrum is removed for each detector independently. Again, this is 
based on a GEANT4 simulation of energy depostion of external 351.9 keV gammas 
in a CdZnTe detector, smeared with the measured energy resolution.

An absolute calibration of the
detectors has been performed after the measurements. Here, a
\iso{Cs}{137} source has been used to determine the relative 
efficiency of the detectors. 
Normalising to the most efficient detector, the remaining two had an efficiency of
67 \% and 80 \% respectively \cite{kie05}. 
The major additional uncertainty is the active volume. However, various information
about the detectors
provided by the supplier and the observation of alpha particles created at the surface 
lead to a conservative upper limit of regions with reduced signals of 10\,\%.  


Finally, the three individual spectra are summed taking the efficiency
of the three detectors into account.  The fit interval ranges from
120\,keV $-$ 310\,keV, the endpoint of the expected spectrum using
the latest number on atomic masses is $Q=320\pm3$\,keV~\cite{aud03}.
The fit range contains a total of about 37000 events, 12.6 \% can be
ascribed to background.

Systematic effects have been checked by varying the ranges of the
fits for the signal and the background regions,
as well as the energy resolution 
which has been folded with the theoretical spectrum. 
All these variations cause an effect
on the calculated half-life of less than 1\%. The slight difference in spectral 
shape between
the two best fit spectra of \cite{dan96} and \cite{ale94} is negligible, being washed
out by the energy resolution. As the typical Zn admixture is between 7-11 \%, the uncertainty in the amount 
of \iso{Cd}{113} in the detector material is 1.8 \%.

\begin{figure}
\centering
\includegraphics[width=\linewidth]{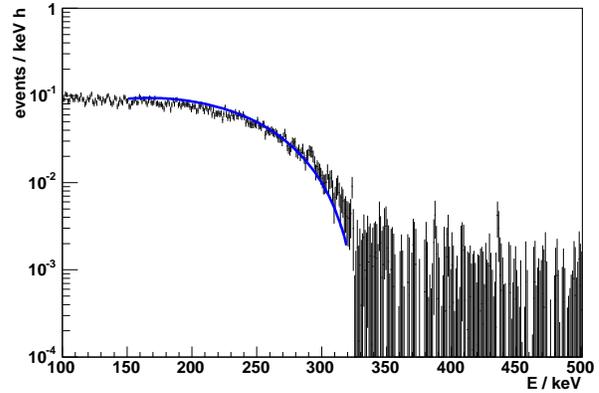}
\caption{The summed low energy spectrum of three CZT detectors after background subtraction. 
The shoulder of \iso{Cd}{113} is the most
prominent feature in that energy range. Also shown is the best spectrum of the type given in \protect \cite{dan96}
describing the data. Note that this is not a fit to the data, only the normalisation was varied.}
\label{sumspectrum}
\end{figure}

The combined energy spectrum is shown in Fig.~\ref{sumspectrum}.
Using the spectral shape of \cite{dan96} including their best fit values results in a decay rate of $(15.98 
\pm 0.41)$ per hour and thus a half-life of
\be
T_{1/2} = (8.2 \pm 0.2 (\rm{stat.}) ^{+0.2}_{-1.0} (\rm{sys.})) \cdot 10^{15} \quad \rm{yrs} \quad .
\ee
This half-life is in good agreement with the values quoted in \cite{ale94,dan96}.
Note that the half-life has not been determined by a real fit to the spectral shape, only the normalisation of the fixed
spectral shape has been changed.
The individual numbers including only statistical uncertainties are $T_{1/2} = (8.1 \pm 0.2) \cdot 10^{15}$ yrs,
$\chi^2/\mathrm{ndf} = 117.2/169$ (Detector 1), $T_{1/2} = (7.5 \pm 0.3) \cdot 10^{15}$ yrs, $\chi^2/\mathrm{ndf} = 160.4/169$ (Detector 3) and 
$T_{1/2} = (9.3 \pm 0.4) \cdot 10^{15}$ yrs, $\chi^2/\mathrm{ndf} = 318.1/169$ (Detector 4).

\begin{figure}
\centering
\includegraphics[width=\linewidth]{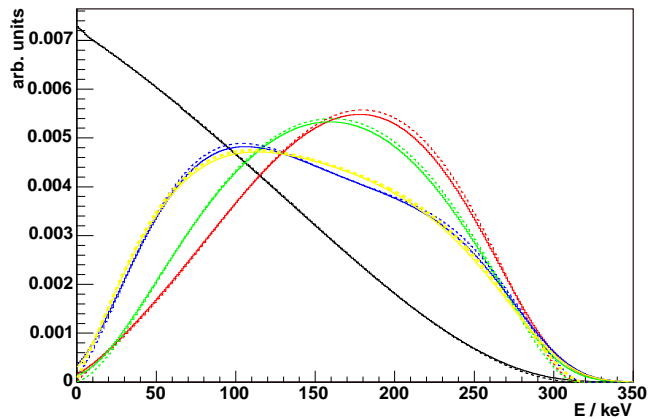}
\caption{Various possible spectra discussed for \iso{Cd}{113} decay. Shown are the spectra of \protect
\cite{dan96} (red), \protect \cite{ale94} (green), a rescaled \iso{In}{115} spectrum \protect \cite{privat} (blue),
the Table of Isotopes Web-page spectrum \protect \cite{toiweb} (black) and the measured spectrum described in this paper (yellow) before applying a cut on the noise threshold. The solid curves correspond to the original spectral shapes, while the dashed lines include the energy smearing due to the energy resolution.}
\label{theospectrum}
\end{figure}


\subsection{Spectral shape of the ${\mathbf{{}^\mathbf{113}}\mathbf{Cd}}$ beta spectrum}
\begin{figure}
\centering
\includegraphics[width=\linewidth]{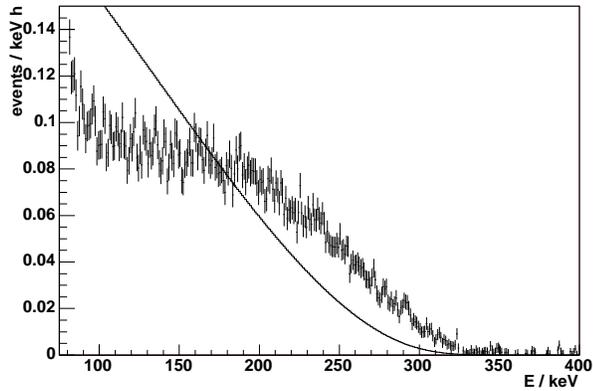}
\caption{Comparison of the observed energy spectrum with the spectral shape 
taken from the Table of Isotopes Web-page \protect \cite{toiweb}, assuming
a Q-value of 320 keV.}
\label{comparetoi}
\end{figure}

Trying a real fit to the data in order to determine a half-life suffers from a number of problems: Several options exist 
to describe a 4-fold forbidden non-unique beta decay spectrum. They are compiled in Fig.~\ref{theospectrum} and 
are already smeared with the energy resolution.
The spectra of \cite{dan96} and \cite{ale94} agree reasonably well. Another option is to compare the spectral shape with the one obtained
for the 4-fold forbidden non-unique \bdec of \iso{In}{115} and to rescale
it to the Q-value of \iso{Cd}{113}. In \cite{pfe79} a polynomial is introduced to describe 
the \iso{In}{115} spectrum. During the development of a low-energy solar neutrino detector based on 
an In-loaded scintillator (LENS) the beta spectrum of \iso{In}{115} has been remeasured and their
best fit values \cite{privat} 
were used to obtain the spectrum shown in Fig.~\ref{theospectrum}. Last, but not 
least, the spectrum from the Table of Isotopes Web-page can be used 
\cite{toiweb}.\\
The measured sum spectrum obtained with the CdZnTe detectors agrees reasonably well with the one expected
from \iso{In}{115}. However, it would have a significant effect on the event number and thus the half-life,
if extrapolated below 120 keV, because the maximum is below 100 keV. 
Additionally, in this energy region  the noise threshold for
the measurement has to be considered. Approximating the threshold by an exponential function and subtracting
it from the total spectrum results in a residual which looks more like the spectrum from 
\cite{dan96} and \cite{ale94}. Thus, it was decided to use their spectral shape, because they have 
already performed measurements of the \iso{Cd}{113} spectrum below 100 keV. 
To solve this issue of different spectral shapes completely, the spectrum has to be measured 
down to at least 50 keV, which will be possible with the next step of COBRA, consisting of 64 detectors.\\
In addition to the polynomial approach, the
electron energy spectrum of the Table of Isotopes Web-page has
been applied \cite{toiweb}. The efficiency correction and energy
folding is done analogously to the procedure described before. 
This spectrum approaches zero only in the lowest energy bin.
As can be seen in Fig.~\ref{comparetoi} the spectral shape is not 
reproduced in the data. The spectra are normalised to the total 
number of events in the range of 100-320 keV.


\section{Summary}
A new double beta decay experiment COBRA is planned using CdZnTe
semiconductor detectors.  As a side product of the test measurements for this project, 
the 4-fold forbidden non-unique
\bdec of \iso{Cd}{113} has been measured.  A first observation using
four crystals clearly shows the signal, but only three of them could be used
for a reliable half-life determination. 
The half-life measurements obtained from each of the detectors are in good agreement with each other and with published values. In the near
future 64 CdZnTe detectors will be running, allowing a precision measurement and systematic
studies of this decay by extracting single independent values for each detector.
This will also include lower noise thresholds allowing the uncertainty which arises from the lack of
knowledge of the spectral shape to be removed.

\section{Acknowledgement}
K.~Zuber is supported by a Heisenberg-Fellowship of the Deutsche Forschungsgemeinschaft.
We thank R.~Firestone and J.~Suhonen for useful discussions and comments.
We are grateful to the LENS collaboration, especially S.~Schoenert and L.~Pandola for providing their new \iso{In}{115} spectrum parameters. 
In addition, we thank the Forschungszentrum Karlsruhe, especially
K.~Eitel, for providing the material for the neutron shield. We thank the mechanical workshop of
the University Dortmund for their support and the Laboratori Nazionali del Gran Sasso (LNGS) for offering
the possibility to perform measurements underground.

\bibliography{cd113}

\begin{thebibliography}{11}
\expandafter\ifx\csname natexlab\endcsname\relax\def\natexlab#1{#1}\fi
\expandafter\ifx\csname bibnamefont\endcsname\relax
  \def\bibnamefont#1{#1}\fi
\expandafter\ifx\csname bibfnamefont\endcsname\relax
  \def\bibfnamefont#1{#1}\fi
\expandafter\ifx\csname citenamefont\endcsname\relax
  \def\citenamefont#1{#1}\fi
\expandafter\ifx\csname url\endcsname\relax
  \def\url#1{\texttt{#1}}\fi
\expandafter\ifx\csname urlprefix\endcsname\relax\def\urlprefix{URL }\fi
\providecommand{\bibinfo}[2]{#2}
\providecommand{\eprint}[2][]{\url{#2}}

\bibitem[{\citenamefont{{Pfeiffer} et~al.}(1979)}]{pfe79}
\bibinfo{author}{\bibfnamefont{L.}~\bibnamefont{{Pfeiffer}}}
  \bibnamefont{et~al.}, \bibinfo{journal}{Phys. Rev. C}
  \textbf{\bibinfo{volume}{19}}, \bibinfo{pages}{1035} (\bibinfo{year}{1979}).

\bibitem[{\citenamefont{{Barabanov} et~al.}(2004)}]{bar04}
\bibinfo{author}{\bibfnamefont{I.}~\bibnamefont{{Barabanov}}}
  \bibnamefont{et~al.}, \bibinfo{journal}{Poster at Neutrino 2004}
  (\bibinfo{year}{2004}).

\bibitem[{\citenamefont{{Singh} et~al.}(1998)}]{sin98}
\bibinfo{author}{\bibfnamefont{B.}~\bibnamefont{{Singh}}} \bibnamefont{et~al.},
  \bibinfo{journal}{Nucl. Data Sheets} \textbf{\bibinfo{volume}{84}},
  \bibinfo{pages}{487} (\bibinfo{year}{1998}).

\bibitem[{\citenamefont{{Mitchell} and {Fischer}}(1988)}]{mit88}
\bibinfo{author}{\bibfnamefont{L.}~\bibnamefont{{Mitchell}}} \bibnamefont{and}
  \bibinfo{author}{\bibfnamefont{P.}~\bibnamefont{{Fischer}}},
  \bibinfo{journal}{Phys. Rev. D} \textbf{\bibinfo{volume}{38}},
  \bibinfo{pages}{895} (\bibinfo{year}{1988}).

\bibitem[{\citenamefont{{Danevich} et~al.}(1996)}]{dan96}
\bibinfo{author}{\bibfnamefont{F.}~\bibnamefont{{Danevich}}}
  \bibnamefont{et~al.}, \bibinfo{journal}{Phys. Atom. Nucl.}
  \textbf{\bibinfo{volume}{59}}, \bibinfo{pages}{1} (\bibinfo{year}{1996}).

\bibitem[{\citenamefont{{Alessandrello} et~al.}(1994)}]{ale94}
\bibinfo{author}{\bibfnamefont{A.}~\bibnamefont{{Alessandrello}}}
  \bibnamefont{et~al.}, \bibinfo{journal}{Nucl. Phys. B Proc. Suppl.}
  \textbf{\bibinfo{volume}{35}}, \bibinfo{pages}{394} (\bibinfo{year}{1994}).

\bibitem[{\citenamefont{Zuber}(2001)}]{zub01}
\bibinfo{author}{\bibfnamefont{K.}~\bibnamefont{Zuber}},
  \bibinfo{journal}{Phys. Lett.B} \textbf{\bibinfo{volume}{519}},
  \bibinfo{pages}{1} (\bibinfo{year}{2001}).

\bibitem[{\citenamefont{{Kiel}}(2005)}]{kie05}
\bibinfo{author}{\bibfnamefont{H.}~\bibnamefont{{Kiel}}},
  \bibinfo{journal}{http://hdl.handle.net/2003/21509}  (\bibinfo{year}{2005}).

\bibitem[{\citenamefont{{Audi} et~al.}(2003)}]{aud03}
\bibinfo{author}{\bibfnamefont{G.}~\bibnamefont{{Audi}}} \bibnamefont{et~al.},
  \bibinfo{journal}{Nucl. Phys. A} \textbf{\bibinfo{volume}{729}},
  \bibinfo{pages}{337} (\bibinfo{year}{2003}).

\bibitem[{\citenamefont{{Schoenert} and {Pandola}}(2004)}]{privat}
\bibinfo{author}{\bibfnamefont{S.}~\bibnamefont{{Schoenert}}} \bibnamefont{and}
  \bibinfo{author}{\bibfnamefont{L.}~\bibnamefont{{Pandola}}},
  \bibinfo{journal}{private communication}  (\bibinfo{year}{2004}).

\bibitem[{\citenamefont{http://ie.lbl.gov}(2004)}]{toiweb}
\bibinfo{author}{\bibnamefont{http://ie.lbl.gov}} (\bibinfo{year}{2004}).

\end{thebibliography}

\end{document}